\documentclass[%
preprint, 
bibnotes,
 amsmath,amssymb,
 aps, physrev,
]{revtex4-2}

\usepackage{braket}
\usepackage{subfigure}
\usepackage{graphicx}
\usepackage{dcolumn}
\usepackage{bm}
\usepackage{hyperref}
\usepackage[monochrome]{xcolor}

\begin{document}

\preprint{}

\title{\textbf{Scattering of Capillary Waves By a Cylindrical Meniscus} 
}%

\author{Cade Sbrocco}
\affiliation{Department of Physics, Cornell University, Ithaca, New York, 14853}
\author{Yukun  Sun}%
\affiliation{Department of Biological and Environmental Engineering, Cornell University, Ithaca, NY 14853 
}%
\author{Uma Grover}%
\affiliation{Department of Biological and Environmental Engineering, Cornell University, Ithaca, NY 14853 
}%
\author{Chris Roh}%
\email{Contact author: cr296@cornell.edu}
\affiliation{Department of Biological and Environmental Engineering, Cornell University, Ithaca, NY 14853 
}%

\begin{abstract}
The propagation of water waves is altered when interacting with curved surfaces. Here, we consider the problem of capillary waves interacting with a 3D cylindrical meniscus. We show that when capillary waves scatter off an object surrounded by a meniscus, the resulting wavefield can be drastically altered by refraction to increase wave presence in the otherwise shadowed region behind a cylinder. This indicates that a meniscus acts as an effective coating around an object, slowing down the wave's propagation speed in its region of influence. Our results are not only an important step in surface tension dominated wave interactions, but may have implications in the biological communication of surface dwelling animals.
\end{abstract}

\maketitle

\section{Introduction} \label{Section:Introduction}
The interactions between capillary waves and menisci are fundamentally important to understanding the dynamics of small objects near the capillary length scale ($\ell_c=2.71$ mm). In nature, surface dwelling organisms may use capillary waves to communicate \cite{bleckmann1985perception} while abiological floating particles can be propelled and interact through waves in the laboratory \cite{harris2025propulsion}. Studies into a meniscus's influence on the scattering of these waves can provide then new insights on the underlying dynamics at play in these systems.

Despite their direct relationship through surface curvature, there are only a handful of studies that directly investigate the scattering interactions between a meniscus and capillary waves. In 2016, experimental studies investigated how the reflection coefficients depend on the height of pinned menisci at a flat wall \cite{michel2016acoustic}. In 2025, complementary experimental and computational studies expanded to considerations of both reflection and transmission coefficients for menisci at shallow, 2D boundaries \cite{wang2025meniscus,liu2025theory}. These studies were limited to 2D effects, and thus, full 3D scattering pattern incurred by the effect of a meniscus remains to be seen.

The archetypal 3D scattering problem for water waves is that of plane waves incident upon a circular cylinder (see Supplemental Material for analytical solution \cite{supp}). This problem is generalizable and has been studied in many other wave systems including water surface gravity waves, electromagnetic waves, and acoustic waves, among others. One may alter the scattering with changes in the properties of the cylinder or additional (effective) coatings. For example, with an appropriate depth profile, one may cloak objects from ocean waves \cite{porter2014cloaking} and tsunamis are naturally focused by the sloping beaches around islands \cite{berry2007focused}. For waves at the capillary-gravity scale, surfactant regimes affect the scattering profiles \cite{chou1995capillary} and electrostriction of the surface allows for tuneable lenses \cite{mouet2023comprehensive}.

\begin{figure}[t]
\includegraphics[width=10cm]{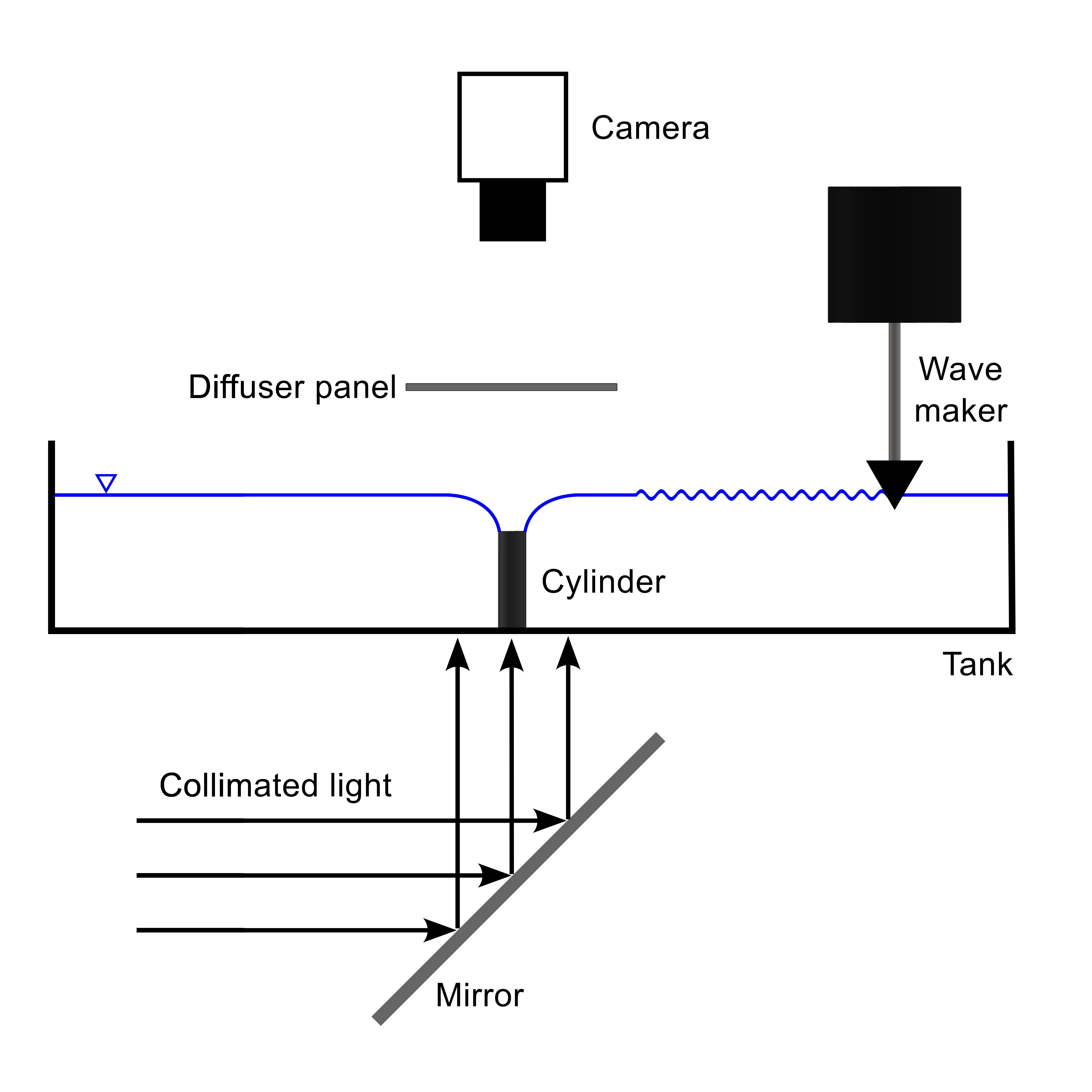}
\caption{Schematic of the experimental setup used to obtain shadowgraphs of capillary waves scattering off of a meniscus.}
\centering
\label{Figure:ExperimentalSetup}
\end{figure}

In this study, we present the scattering of capillary waves by a 3D meniscus formed around an immersed vertical cylinder. We performed experiments in a wave tank using free-surface shadowgraphy (Fig. \ref{Figure:ExperimentalSetup}) to obtain the resulting wavefield amplitudes. The results show that a meniscus alters the scattering pattern by acting as an effective media coating an object, similar to the sloping beaches around islands in gravity dominated systems. Both positive and negative menisci are observed to refract the waves to varying degree into the otherwise shadowed region, indicating an on average decrease in wave speed.

\section{Experimental Setup}
A $60$ cm x $36$ cm rectangular container was filled with water to a depth of $30$ mm. A plunger-style wave maker spanning most of the width of the tank and many wavelengths away from the cylinder was attached to a magnetic shaker (B\&K Type 4810). This was connected to a signal generator (Tektronix AFG3101C) in tandem with a power amplifier (B\&K Type 2706) to generate continuous, small amplitude, monochromatic capillary waves that were deemed sufficiently planar. The cylinders used as scattering objects were 3D printed out of resin to around $0.05$ mm accuracy. For negative menisci, the tops were coated with superhydrophobic paint (NeverWet). For positive menisci, a plasma cleaner (Harrick PDC-32G) was used to activate the surface of the cylinders to decrease the contact angle. The contact line was fixed by pinning the surface to the rim of the cylinder (see Supplemental Material \cite{supp}).

We performed shadowgraph measurements by shining collimated light from below the tank with the refracted light pattern rear projected onto white diffusing glass (near-Lambertian) a distance $45$ mm above the water. A high speed camera (Photron FASTCAM mini AX 200) was positioned above the viewing plate to record the pattern at $750$ fps with a resolution close to $9.5$ pixels per millimeter. When menisci were present, light rays within a distance of $5\ell_c$ of the origin were blocked as shadowgraphs are a linear approximation in regions of mild curvature (see Supplemental Material for more details \cite{supp}). The videos were average subtracted and each pixel was fit to a sinusoid function to obtain intensity amplitudes.

\section{Problem Statement}
In our otherwise quiescent air-water system, the surface tension $\sigma$, fluid density $\rho$, gravitational acceleration $g$, and viscosity $\mu$ are fixed. A cylinder of radius $a$ with a static (under quiescent conditions) contact angle $\varphi_c$ uniquely determines a circularly symmetric meniscus profile (with a unique contact line height $h_c$). With a solid, semi-infinite cylinder (opaque), as opposed to a thin disc (translucent), we do not need to consider finite depth or transmission effects. Additionally, pinning the meniscus to the rim of the cylinder allows us to avoid considerations of slip effects \cite{mahdmina1990scattering,zhang2013capillary}. Finally, the incident waves are to be monochromatic (with frequency $f$, wavenumber $k$) to avoid dispersive effects, sufficiently within the capillary regime to neglect gravitational effects, planar, and are of small characteristic amplitude $\epsilon$ to avoid non-linear effects away from the meniscus \cite{crapper1957exact}. 

Our system then reduces to five non-dimensional parameters, of which only the first three are effectively tuneable: the static contact angle $\varphi_c$, the radius $\frac{a}{\ell_c}$, the scattering size parameter $X=ka$, the small characteristic amplitude $\epsilon k \ll 1$, and the fixed Laplace number $\text{La} = \frac{\sigma \rho \ell_c}{\mu^2}$ \footnote{One can equivalently non-dimensionalize as to instead obtain the capillary number $\text{Ca} = \frac{\mu c_g}{\sigma}$, where $c_g$ is the group velocity. While this is not a constant of the system, one may view this as physically more meaningful than the Laplace number in some contexts since a crude estimate of the rate of decay of capillary waves is $\gamma = 2\text{Ca}k$.}, where $\ell_c = \sqrt{\frac{\sigma}{\rho g}}=2.71$ mm is the capillary length which will be used as our length scale in all discussions and plots going forward ($\ell_c=1$).

\section{Results and Discussion}\label{Section:Results and Discussion}
\subsection{Experimental Results}
\begin{figure*}[t]
\includegraphics[width=14.5cm]{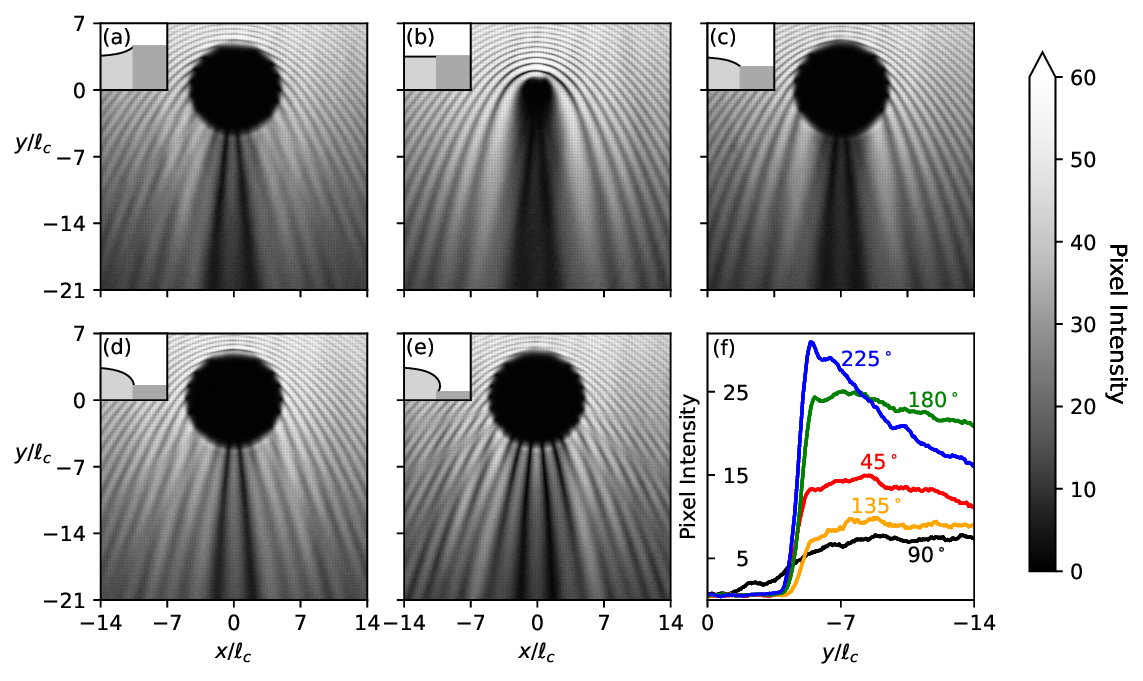}
\caption{(a-e) Shadowgraph intensity amplitude fields for menisci with static contact angles $\varphi_c=45^\circ,90^\circ,135^\circ,180^\circ,225^\circ$, respectively. For monochromatic waves, the pixel intensity is sensitive to the wave height \cite{supp}. The corresponding meniscus profiles (bold black line) are shown in the inset figures (not to scale). All setups use the same cylinder radius, but for figures (a,c-e) light rays in the shadowgraph incident within a distance $r=5\ell_c$ of the origin were blocked to avoid refracted light in regions of large curvature due to the linear approximation of shadowgraphs \cite{supp}. Pixel intensity values are directly comparable between figures. (f) Intensity profiles along the centerline of the cylinder. A low pass filter was used to remove noise.}
\centering
\label{Figure:ShadowgraphAmplitude}
\end{figure*}
We obtained shadowgraphs for $k\approx \frac{2\pi}{\ell_c}$ ($f=150$ Hz) waves scattering off of cylinders with varying static contact angle and of radius $a=2 \ell_c$ (scattering size parameter $X=4\pi$). We projected the 3D system onto a 2D screen to obtain images where the pixel intensity of the wave field outside the meniscus which is sensitive to the surface height for small amplitude, monochromatic waves (see Supplemental Material \cite{supp}). Each pixel's intensity amplitude was found to obtain the total wavefield amplitudes shown in Figs. \ref{Figure:ShadowgraphAmplitude}(a-e). To aid in the visualization of discussed phenomena, the image contrast was increased uniformly across the figures.

\textit{Flat Meniscus} ($\varphi_c=90^\circ$), Fig. \ref{Figure:ShadowgraphAmplitude}(b): When no meniscus is present the amplitude field bares a strong resemblance to the expected ideal scattering solution (see Fig. \ref{Figure:Diffraction}(b) and Section \ref{Section:Interpretations}), albeit, with some viscous decay. As expected, a mostly shadowed region with minimal illumination (increased wave amplitude) solely due to diffraction is observed directly behind the cylinder with intensity maxima and minima due to interference between the incident and scattered waves (see Section \ref{Section:Interpretations} for more discussion). We use this as our base case for further comparisons as any change to the amplitude cannot then be solely attributed to diffractive effects of the cylinder alone.

\textit{Positive Meniscus} ($\varphi_c=45^\circ$), Fig. \ref{Figure:ShadowgraphAmplitude}(a): If the cylinder is above the mean water level, a positive meniscus forms. As water waves scatter off of the meniscus, we observe a noticeable change in the amplitude field behind the cylinder. The meniscus appears to refract the waves to behind the cylinder, illuminating the previously shadowed region with a higher intensity central maximum. Additionally, what was initially a wide, first intensity maximum in the flat meniscus case (noticeable approximately between $|x|\in[2,7]$ along the $y=-21$ in Fig. \ref{Figure:ShadowgraphAmplitude}(a)) has now begun to split noticeably into two maxima (with minima at $|x|\approx 6$).

\textit{Negative Meniscus} ($\varphi_c=135^\circ$), Fig. \ref{Figure:ShadowgraphAmplitude}(c): If the cylinder is below the mean water level, a negative meniscus is formed. At the chosen static contact angle, the meniscus profile is complementary to the previous positive meniscus (the surface is symmetrically mirrored across the mean water level). In this case the amplitude field exhibits different behavior to its compliment, showing only a subtle changes when compared to the flat meniscus with a slightly more illuminated shadowed region.

\textit{Extreme Negative Menisci} ($\varphi_c=180^\circ,225^\circ$), Figs. \ref{Figure:ShadowgraphAmplitude}(d,e): We further increase the static contact angle resulting in negative menisci which vertically meet the cylinder, and even overturn. Even starker changes in the amplitude field are seen with the furthering increase of the contact angle resulting in more complicated and contrasted features. Perhaps most noticeably, the originally shadowed region is even more illuminated. Including the center, we count $11$ distinct intensity maxima along the $y=-21$ in the $\varphi_c=225^\circ$ meniscus, in comparison to only $7$ for the flat meniscus case in Fig. \ref{Figure:ShadowgraphAmplitude}(b). Lastly of note, is the relative intensity between maxima in the $\varphi_c=225^\circ$ meniscus, with the second intensity maxima from the center being considerably brighter than the others.

To quantify the relative degree of illumination of the shadowed region between the cases above, the pixel intensity profiles through the centerline ($x=0$) of the cylinders are show in Fig. \ref{Figure:ShadowgraphAmplitude}(f). The mild negative meniscus ($\varphi_c=135^\circ)$ does in fact slightly illuminate the shadowed region more than flat meniscus, but not to the same degree as its complimentary positive meniscus ($\varphi_c=45^\circ$) or the extreme negative menisci. While refractive phenomena may be usually thought of in the context of the ray optics regime ($X\gg1$), it is additionally observed at moderate scattering size parameters in which simulations of simple coatings can lead to qualitatively similar wavefields as those observed in our experiments (see Supplemental Material \cite{supp}). As the wavefronts appear to bend around to behind the object, the menisci must be slowing down the propagation in their region of influence.

Of special note in Fig. \ref{Figure:ShadowgraphAmplitude}(f) is the intensity profile of the extreme $\varphi_c=225^\circ$ negative meniscus. Compared to the others, the observed amplitude profile not only has the highest amplitude maximum, but it decreases at a much faster rate. We hypothesize that this indicates lensing behavior in which a focal region may be present. Lensing of water waves is seen both in capillary waves through electrostriction \cite{mouet2023comprehensive} and larger, gravity dominated water waves due to varying depth profiles \cite{berry2007focused}, to which we liken our system as a capillary analogue. While a focal region may be present in this system, our current experimental methods preclude its observation since it would be on the meniscus.

\subsection{Interpretations}\label{Section:Interpretations}
In the flat meniscus case, the observed wavefield should be governed effectively entirely by diffractive effects from the prescribed boundary conditions at the cylinder. Any differences in the scattering then must be the result of refractive effects due to a meniscus affecting the wave propagation. While analytic solutions can found by singly applying a no-penetration / fully slipping ($\phi_r(a) = 0$) boundary condition (see Supplemental Material \cite{supp}), attempting to additionally enforce pinning leads to difficulties \cite{mahdmina1990scattering,zhang2013capillary}. In this case an additional short range force is expected to act upon the waves leading to significant increases in the wave speed of capillary-gravity waves in waveguides \cite{monsalve2022space,scott1978waves} and arrays of objects \cite{fauconnier2025fast} compared to non-pinned systems.
\begin{figure*}[t]
\includegraphics[width=16cm]{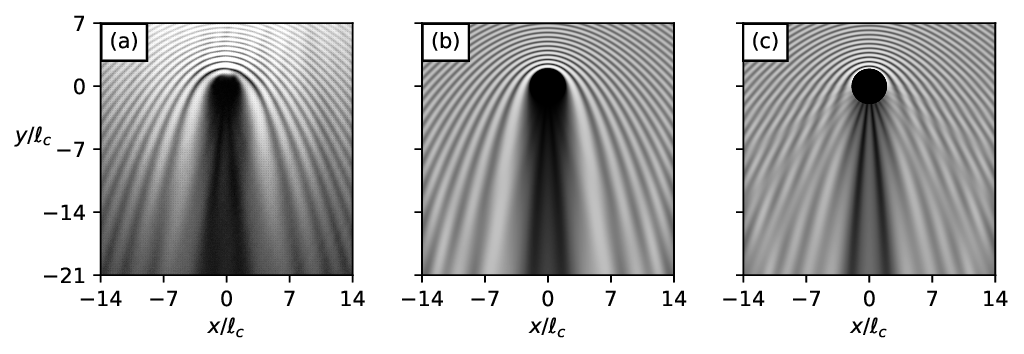}
\caption{Comparison between the (a) experimental flat meniscus result and equivalent ideal scattering simulations with (b) Dirichlet (pinned) and (c) Neumann (no penetration / fully slipping) boundary conditions. The simulations were performed in the FDTD solver MEEP \cite{oskooi2010meep} with a matched scattering size parameter $X=4\pi$.}
\centering
\label{Figure:Diffraction}
\end{figure*}

To investigate how these two boundary conditions simultaneously act on our system, we compare our flat meniscus result to ideal wave scattering simulations (linear, non-dissipative), one with a Dirichlet (pinned, $\eta(a)=0$) \footnote{In an air-water system, solely applying the Dirichlet boundary condition is nonphysical as a semi-infinite cylinder will still require the no-penetration boundary condition while a thin disc will allow for flow underneath.} and one with a Neumann (no-penetration / fully slipping) boundary condition (Fig. \ref{Figure:Diffraction}(b,c)). Matching the scattering size parameter $X=4\pi$, we notice that qualitatively, a Dirichlet boundary condition yields a strikingly similar wavefield to our experimental result for the flat meniscus (Fig. \ref{Figure:Diffraction}(a)). While the Neumann boundary condition may qualitatively appear somewhat similar to our meniscus results, especially the $\varphi_c=180^\circ$ meniscus, this is only superficial as our contact lines are pinned, and so this must be due to different mechanisms. Lastly, there appears to be no prominent effects indicative of non-negligible increased wave speed expected from higher-order pinning effects \cite{monsalve2022space,scott1978waves}. Thus, the resulting wavefield from scattering waves off of a single cylinder with a flat, pinned meniscus is sufficiently described by diffraction due to the pinned contact line in our setup.

\begin{figure*}[t]
\includegraphics[width=16cm]{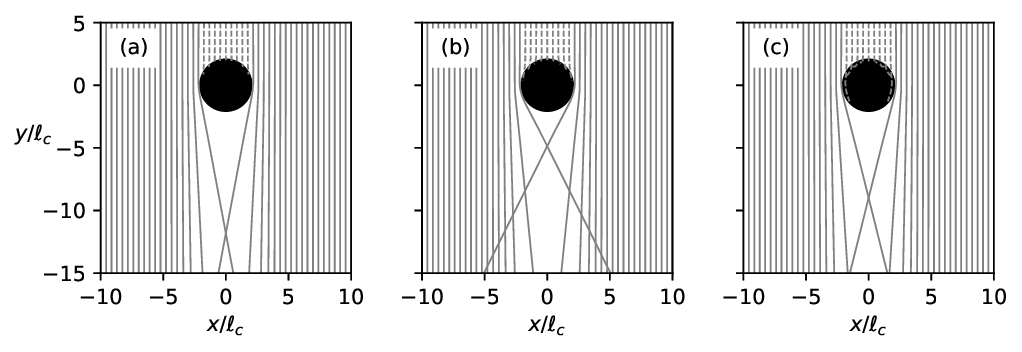}
\caption{Geodesics on (a) $45^\circ/135^\circ$, (b) $0^\circ/180^\circ$, and (c) $-45^\circ/225^\circ$ menisci for initially planar rays. Rays which will reflect off of the cylinder are denoted by dashed lines and truncated.}
\centering
\label{Figure:Geodesics}
\end{figure*}

Next, to begin identifying the possible mechanisms behind the refractive phenomena observed when a meniscus is present, we first limit our consideration to waves in the ray limit confined to propagate along a surface $S$. If the surface appears sufficiently planar such that the dispersion relation remains unchanged across the surface, the ray trajectories must be the geodesics of the surface. While the absolute speed of the wave is unchanging, its projected speed in the $xy$-plane will appear to be slowed (what would be observed in our shadowgraphs). This in itself can result in lensing phenomena, known in the context of gravitational lensing \cite{walsh19790957+} and optical devices  \cite{righini1973geodesic}. We can additionally observe lensing of rays on meniscus surfaces, strongly refracting glancing and impinging rays as seen in Fig. \ref{Figure:Geodesics}.

These effects are not dependent on the sign of the surface curvature, and thus will be the same for complimentary positive and negative menisci. As such, the asymmetry in our results precludes significant attribution to geodesic effects, at least in the mild menisci. Furthermore, in practice, pure geodesic trajectories will not be observed in our system, obscured by refractive and diffractive effects. Nevertheless, the waves are still confined to surface and so these effects may be of more importance near the contact line of extreme menisci.

We now look to identify those physical mechanisms which will affect the dispersion relation (absolute speed) of the waves. Let $S$ explicitly be an air-water interface bounding an inviscid fluid of volume $V$ (see Section \ref{Section:AdditionalMechanisms} for more discussion on viscosity). The dynamic boundary condition of the surface for a small perturbation $\eta$ is
\begin{equation} \label{Equation:Dynamic}
    \rho \phi_t =-\sigma \kappa,\text{ on } S
\end{equation}
where $\phi$ is the velocity potential and $\kappa$ is the mean curvature of the perturbed surface. This equation represents the conservation of energy between the inertia and surface tension. The velocity potential due to the geometry is determined by the incompressibility condition
\begin{equation} \label{Equation:Incompressibility}
    \nabla^2\phi = 0,\text{ in } V
\end{equation}
and the system of equations is complete with the kinematic boundary condition
\begin{equation} \label{Equation:Kinematic}
    \eta_t = \hat{n}\cdot\nabla\phi,\text{ on } S
\end{equation}
in addition to any other prescribed boundary conditions. All together, the interplay between the surface tension and inertial effects will determine the wave dynamics. For an infinitely deep system with an otherwise flat surface, the dispersion relation for a wave is
\begin{equation} \label{Equation:FlatDispersion}
    c_0=\sqrt{\frac{\sigma}{\rho}k} =\frac{\text{Surface Tension Effects}}{\text{Inertial Effects}}
\end{equation}
where $c_0$ is the phase speed. In cases where there is a curved surface geometry or depth profile, the dispersion relation will be modified in such a way that we can define an index of refraction throughout
\begin{equation} \label{Equation:IndexOfRefraction}
    N_\text{disp}=\frac{c_0}{c}=\frac{\text{Inertial Effect Modification}}{\text{Surface Tension Effect Modification}}
\end{equation}
Only in simple geometries can solutions be analytically found \cite{rayleigh1878instability}, of which our system is not one \cite{supp}. In general, $N_\text{disp}$ can be both greater than or less than unity, complex, inhomogeneous, frequency dependent, and anisotropic. 

\textit{Modification to Surface Tension Effects}: The meniscus and perturbation profiles contribute non-linearly to the surface curvature. Further complexity emerges when the curvature varies sufficiently over a wavelength. These effects are expected to be most relevant in the strongly curved regions near the contact line in extreme menisci. Furthermore, as these effects are independent of the sign of surface curvature, they will affect positive and negative menisci symmetrically.

\textit{Modification to Inertial Effects}: The change in the geometry of a curved surface system will affect how a wave's influence penetrates into the bulk of the fluid. This stems from both the curvature of the surface and, in the case of positive menisci, the finite projected depth between the surface and the cylinder. In contrast to the geodesic and surface tension effects, these effects are dependent on the sign of surface curvature, and thus will affect positive and negative menisci asymmetrically.

It is clear then from the asymmetry between the complementary positive ($\varphi_c=45^\circ$, Fig. \ref{Figure:ShadowgraphAmplitude}(a)) and negative ($\varphi_c=135^\circ$, Fig. \ref{Figure:ShadowgraphAmplitude}(c)) menisci that inertial effects must play an important role in the scattering to result in the observed phenomena. Specifically for mild menisci, this appears to be a dominant mechanism as only the positive meniscus appears to significantly affect the scattering. For the extreme menisci however, the claim of a single important mechanism may be inappropriate and likely requires holistic consideration of inertial, surface tension, and geodesic effects. This conclusion agrees with interpretations of results in previous wave-meniscus interaction studies at lower frequencies ($5-25$ Hz) where larger wavelengths relative to meniscus size lead to increased modifications of surface tension effects \cite{wang2025meniscus}.

\subsection{Additional Mechanisms}\label{Section:AdditionalMechanisms}
In different parameter regimes or geometries, other physical mechanisms beyond those considered above may play prominent roles in altering the scattering behavior.

\textit{Viscous Damping Effects}: While outside the scope of the interpretations in this study, it would be careless to fully disregard the discussion of the contribution of damping effects in our system. While the defined $N_\text{disp}$ in Eq. \ref{Equation:IndexOfRefraction} does not account for viscosity, its effects are increasingly prevalent at higher frequencies. Already the dispersion relation is affected by viscosity on the flat surface for the incident waves \cite{Crapper1984}, and further effects are known to occur from the meniscus and near the boundary \cite{huang2020streaming, kidambi2009meniscus}. Specifically for the scattering of larger capillary-gravity waves, direct measurements of damping effects at pinned menisci were shown to significantly alter the reflection coefficient at flat walls, more so with extreme menisci \cite{michel2016acoustic}. We expect damping effects to be intimately connected with the inertial effects as they both are directly concerned with interactions within the bulk of fluid, and thus, will additionally act asymmetrically on positive and negative menisci. Thus, further characterization of damping effects will be important in future studies of wave-meniscus interactions.

\textit{Gravitational Effects}: As the waves we investigated are of sufficiently high frequency as to be considered pure capillary waves, gravity should effectively only play a role in the formation of the meniscus. However, at lower frequencies, gravity's role as a restoring force grows relative to the surface tension. In addition to modifying the dispersion relation directly for the flat surface, there is the added complexity of the restoring force no longer necessarily being normal to a curved surface \cite{liu2025theory}.

\textit{Convective Effects}: At sufficiently high amplitudes, the capillary waves are no longer able to be considered linear, and the propagation and shape of the waves will be affected \cite{crapper1957exact,punzmann2014generation}. Further effects may appear when significant underlying flow is present, altering the propagation of waves and generating steady and unsteady wakes. As our waves are linear and the system is otherwise quiescent, these convective effects should not be present.

\section{Outlook}\label{Section:Outlook}
In this study, we performed experiments to investigate capillary waves scattering off a meniscus in a 3D system. Our experimental results show that the meniscus's presence acts as an effective coating to objects, leading to phenomena such as increased wave amplitude behind an obstacle. Across positive and negative menisci, we observe asymmetric behavior but a general trend of a decrease in wave speed due to their presence. From this, we conclude that inertial effects play a large role in affecting the scattering for positive mild menisci while geodesic, surface tension, and inertial effects likely play a combined role for more extreme menisci.

Beyond cylindrical geometries, one may design novel meniscus profiles formed at membranes \cite{schrecengost2025shape} and objects of varying shapes \cite{delens20253d} which may allow for increase control over capillary wave propagation. Considerations of menisci could also allow for greater insight into studies of interacting, wave-propelled floating particles \cite{harris2025propulsion}. Lastly, our results are of relevance to biological communication. Small water surface dwelling animals such as whirligig beetles, water striders, and fishing spiders are known to be highly sensitive to capillary-gravity waves \cite{bleckmann1985perception}. Our results suggest that the meniscus around these animals may play the role of an abiological lens, allowing for increased perception. 

\section{Acknowledgments}
This work was supported by National Science Foundation Grants: NSF-CBET-2442036.

\section{Data Availability}
Data is available upon request.

\bibliography{output}
\end{document}